\documentclass[twoside,twocolumn,9pt]{article}
\usepackage{extsizes}
\usepackage[super,sort&compress,comma]{natbib} 
\usepackage[version=3]{mhchem}
\usepackage[left=1.5cm, right=1.5cm, top=1.785cm, bottom=2.0cm]{geometry}
\usepackage{balance}
\usepackage{widetext}
\usepackage{times,mathptmx}
\usepackage{sectsty}
\usepackage{graphicx} 
\usepackage{lastpage}
\usepackage[format=plain,justification=raggedright,singlelinecheck=false,font={stretch=1.125,small,sf},labelfont=bf,labelsep=space]{caption}
\usepackage{float}
\usepackage{fancyhdr}
\usepackage{fnpos}
\usepackage[english]{babel}
\usepackage{array}
\usepackage{droidsans}
\usepackage{charter}
\usepackage[T1]{fontenc}
\usepackage[usenames,dvipsnames]{xcolor}
\usepackage{setspace}
\usepackage[compact]{titlesec}

\usepackage{epstopdf}

\definecolor{cream}{RGB}{222,217,201}

\newcommand{\eg}{\textit{e.g.}$\,$}
\newcommand{\ie}{\textit{i.e.}$\,$}
\newcommand{\iring}{{$I_{\text{ring}}\,$}}
\newcommand{\rr}{\mathbf{1}}

\newcommand{\drr}{d_{\mathbf{1}}}

\begin{document}

\pagestyle{fancy}
\thispagestyle{plain}
\fancypagestyle{plain}{

}

\makeFNbottom
\makeatletter
\renewcommand\LARGE{\@setfontsize\LARGE{15pt}{17}}
\renewcommand\Large{\@setfontsize\Large{12pt}{14}}
\renewcommand\large{\@setfontsize\large{10pt}{12}}
\renewcommand\footnotesize{\@setfontsize\footnotesize{7pt}{10}}
\makeatother

\renewcommand{\thefootnote}{\fnsymbol{footnote}}
\renewcommand\footnoterule{\vspace*{1pt}%
\color{cream}\hrule width 3.5in height 0.4pt \color{black}\vspace*{5pt}} 
\setcounter{secnumdepth}{5}

\makeatletter 
\renewcommand\@biblabel[1]{#1}            
\renewcommand\@makefntext[1]%
{\noindent\makebox[0pt][r]{\@thefnmark\,}#1}
\makeatother 
\renewcommand{\figurename}{\small{Fig.}~}
\sectionfont{\sffamily\Large}
\subsectionfont{\normalsize}
\subsubsectionfont{\bf}
\setstretch{1.125} 
\setlength{\skip\footins}{0.8cm}
\setlength{\footnotesep}{0.25cm}
\setlength{\jot}{10pt}
\titlespacing*{\section}{0pt}{4pt}{4pt}
\titlespacing*{\subsection}{0pt}{15pt}{1pt}

\fancyfoot{}
\renewcommand{\headrulewidth}{0pt} 
\renewcommand{\footrulewidth}{0pt}
\setlength{\arrayrulewidth}{1pt}
\setlength{\columnsep}{6.5mm}
\setlength\bibsep{1pt}

\makeatletter 
\newlength{\figrulesep} 
\setlength{\figrulesep}{0.5\textfloatsep} 

\newcommand{\topfigrule}{\vspace*{-1pt}%
\noindent{\color{cream}\rule[-\figrulesep]{\columnwidth}{1.5pt}} }

\newcommand{\botfigrule}{\vspace*{-2pt}%
\noindent{\color{cream}\rule[\figrulesep]{\columnwidth}{1.5pt}} }

\newcommand{\dblfigrule}{\vspace*{-1pt}%
\noindent{\color{cream}\rule[-\figrulesep]{\textwidth}{1.5pt}} }

\makeatother

\twocolumn[
  \begin{@twocolumnfalse}
\vspace{0.5cm}
\sffamily
\begin{tabular}{m{0.0cm} p{18cm} }

& \noindent\LARGE{\textbf{Electronic Aromaticity Index for Large Rings$^\dag$}} \\
\vspace{0.0cm} & \vspace{0.3cm} \\
&
\noindent\large{\textbf{Eduard Matito$^{\ast}$\textit{$^{a,b,c}$}}}\vspace{0.5cm} \\

& \noindent\normalsize{We introduce a new electronic 
aromaticity index, AV1245, consisting
in the average of the 4-center MCI values along the ring that keep a positional relationship 
of 1,2,4,5. AV1245 measures the extent of transferability of the delocalized electrons between
bonds 1-2 and 4-5, which is expected to be large in conjugated circuits and, therefore, in
aromatic molecules.
A new algorithm for the calculation of MCI for large rings is also introduced and used to
produce the data for the calibration of the new aromaticity index.
AV1245 does not rely on reference values, does not suffer from large numerical precision errors,
and it does not present any limitation on 
the nature of atoms, the molecular geometry or the level of calculation.
It is a size-extensive measure with a small computational cost that grows linearly with
the number of ring members. Therefore, it is specially suitable to study the aromaticity
of large molecular rings as those occurring in belt-shaped M\"obius structures or porphyrins.
} \\

\end{tabular}

 \end{@twocolumnfalse} \vspace{0.6cm}

  ]

\renewcommand*\rmdefault{bch}\normalfont\upshape
\rmfamily
\section*{}
\vspace{-1cm}


\footnotetext{\textit{$^{a}$~Faculty of Chemistry, University of the Basque Country UPV/EHU,
and Donostia International Physics Center (DIPC). 
P.K. 1072, 20080 Donostia, Euskadi, Spain.}}
\footnotetext{\textit{$^{b}$~IKERBASQUE, Basque Foundation for Science, 48011 Bilbao, Spain.}}
\footnotetext{\textit{$^{c}$~Instituto de Qu\'imica at the Universidad Aut\'onoma de M\'exico (UNAM), Mexico.}}





\section{Introduction}

Chemistry is essentially an experimental science that evolved through experimentation and it has been built upon a series of empirically proved laws.
On the other side, quantum mechanics relies on postulates from which a solid theory has been constructed. 
Both focus on the study of matter, however, 
quantum mechanics can anticipate the electronic
structure of matter and it could, in principle, replace the laws and models of chemistry by physically sound theories.
Notwithstanding, after many years of the advent of quantum chemistry, several chemical concepts with high predictable
power still prevail. 
Most of these concepts have
not found (and most likely cannot find) a solid root in the quantum theory because 
there is no observable behind them. One finds many such concepts in the literature
(\eg, chemical bonding, bond order, ionicity, electron population, agostic bond, etc.)
that are still widely used to predict or explain the electronic structure of molecules or reaction
mechanisms.~\cite{frenking:07jcc}
One of the most employed
terms in literature ---and one of the most controversial~\cite{hoffmann:15as} ones---
is \textit{aromaticity}.~\cite{schleyer:96pac,poater:05cr,chen:05cr,krygowski:14cr,feixas:15csr}
Aromaticity is associated with cyclic electron delocalization in closed circuits that gives rise
energy stabilization, bond length equalization, large magnetic anisotropies and abnormal
chemical shifts, among other effects. Various of these aromaticity manifestations can be
measured by appropriate quantities, the aromaticity indices, that allow for aromaticity
scales. As a result, nowadays there is a number of indices available in the literature, 
often offering disparate results about the aromaticity of certain chemical species.~\cite{katritzky:89jacs}
The discovery of new aromatic species~\cite{li:01sci,boldyrev:05cr,islas:11acr}
that extend well beyond the realm of organic chemistry
has challenged our understanding of aromaticity and it has put forward the limitations of 
the existing aromaticity indices to deal with these new chemical creatures.~\cite{feixas:13wir}
We have contributed
to calibrate and test the limits of applicability of these measures by designing a 
series of tests that aromaticity indices should pass.~\cite{feixas:08jcc,feixas:10jctc} 
The indices based on
multicenter electron sharing have been the most successful ones. These indices are free of 
arbitrary references, can be applied to organic and inorganic species alike and, their latest
versions are free of the ring-size dependency of the original definitions.~\cite{cioslowski:07jpca}
Despite these benefits, the multicenter indices suffer from a series of problems that prevents its application
in large rings as those showing 
in new interesting problems such as the aromaticity of (expanded~\cite{alonso:15cej}) 
porphyrins~\cite{feixas:09cjc} and its relationship with UV absorption spectra or
H\"uckel-to-M\"obius topological switches.~\cite{yoon:11jpcb,marcos:12jpcc}
The goal of this paper is introducing a new electronic aromaticity index that can be applied to rings of arbitrary size.
The work is organized as follows: first, we will review the expressions
for multicenter indices and reveal their strenghts and weaknesses; second, we will describe an algorithm that
permits to calculate MCI on larger rings but it is still limited to rings of medium size;  
then we will provide the expression of a
new multicenter index that will be finally analyzed in a series of compounds to assess its performance.

\section{Multicenter Indices}

In the following we will indicate the coordinates of the electron using the 
short-hand notation $\rr\equiv(\vec{r}_1,\sigma_1)$ and $\drr\equiv d\vec{r}_1d\sigma_1$ for the derivatives. 
Unless otherwise indicated, we will assume a wavefunction constructed from a single-determinant, 
for a more general approach
we suggest Ref.~\citenum{feixas:14jctc}.
We will measure the aromaticity in a ring structure that consists of $n$ atoms,
represented by the string
$\mathcal{A}$=\{$A_1$,$A_2$,...,$A_n$\}, whose elements are
ordered according to the connectivity of the atoms in the ring.\newline

To our knowledge, Giambiagi was first to develop an index to measure
multicenter bonding.~\cite{giambiagi:90sc} The same expression applied
to a molecular ring was named \iring and used
to account for aromaticity~\cite{giambiagi:00pccp}
\begin{eqnarray}\label{eq:iring}
I_{\text{ring}}
(\mathcal{A})=2^{n-1}
\sum_{i_1i_2\ldots i_n} S_{i_1i_2}(A_1)\cdots S_{i_ni_1}(A_n)
\end{eqnarray}
where $S_{ij}(A_1)$ is the atomic overlap matrix (AOM) of atom $A_1$, 
\begin{equation}\label{eq:sij}
S_{ij}(A_1)=\int_{A_1} \drr \phi^*_i(\rr)\phi_j(\rr)   \,,
\end{equation}
and $\phi_i(\rr)$ is a molecular spinorbital. 
\iring depends on the order of the atoms in the string for $n>3$.\newline

Bultinck et al.~\cite{bultinck:05jpoc} suggested MCI, which is computed by summing 
up all the possible \iring contributions 
obtained from the permutations of the atoms in $\mathcal{A}$,
\begin{eqnarray}\label{eq:MCI}
\text{MCI}(\mathcal{A})=\frac{1}{2n}\sum_{\mathcal{P(A)}} 
I_{\text{ring}} (\mathcal{A})
\end{eqnarray}
where $\mathcal{P}$ is an operator that acting on $\mathcal{A}$ produces
all the $n!$ permutations of its atoms.
The MCI is related to the $n$-center electron sharing index
($n$c-ESI),~\cite{ponec:97jpca,sannigrahi:99cpl,francisco:07jcp,feixas:14jctc} $\delta(\mathcal{A})$, by 
\begin{eqnarray}\label{eq:ncESI}
\text{MCI}(\mathcal{A})=\frac{(n-1)!}{2}\delta(\mathcal{A})
=(-2)^{n-2}
\left<\prod_{i=1}^{n}\left(\overline{N}_{A_i}-\hat{N}_{A_i}\right)\right>
\end{eqnarray}
with $\hat{N}_A$ being the particle operator applied to atom $A$ 
and $\overline{N}_A$ the average number of electrons in $A$,
\begin{eqnarray}
\bar{N}(A)=\int \hat{N}_A \rho(\rr)d_{\rr} \equiv \int_A \rho(\rr)d_{\rr}
\end{eqnarray}

The r.h.s. of Eq.~\ref{eq:ncESI} indicates that $\text{MCI}(\mathcal{A})$ is a mesure of how 
the electron distribution of $n$ particules in $n$ atoms $A_1,\ldots,A_n$
is skewed from its mean and it is related to the simultaneous electron fluctuation
among these atomic basins. Therefore, \iring$\,$ measures the delocalization \textit{along} the ring
while MCI also accounts for the delocalization \textit{across} the ring. In the author's
opinion \iring$\,$ should be connected to \textit{cyclic electron delocalization}, which is the term
used by Schleyer in the IUPAC to define aromaticity.~\cite{schleyer:96pac} 
On the other hand, MCI measures the global electron
delocalization that is obviously not limited to the cyclic electron delocalization along the
ring structure. This distinction becomes important for small rings (typically four- and
five-membered rings) where cross-contributions of the electron delocalization might be more 
important.~\cite{castro:14cej}
\newline

These expressions, Eqs.~\ref{eq:iring} and~\ref{eq:MCI}, suffer from ring-size
dependency. Indeed, the overlaps entering Eq.~\ref{eq:sij} are, in absolute value, usually smaller than zero
and, therefore, the larger the ring the smaller \iring and MCI values. 
A few years ago, we suggested a
normalization that not only avoids this problem but it also closely matches the classical topological
resonances per $\pi$ electron (TREPE)~\cite{gutman:77jacs} 
of aromatic annulenes and their ions.~\cite{cioslowski:07jpca}
In this paper, for the sake of simplicity, we just retain the normalization that produces
ring-size-independent indices by taking the $1/n$ power of Eqs.~\ref{eq:iring} and~\ref{eq:MCI}, 
\ie, $I_{\text{ring}}^{1/n}$ and $MCI^{1/n}$. If the value is negative, we will take the
power of the absolute value and multiply the resulting number by minus one.

\section{Multicenter Indices Drawbacks}
\subsection{The atomic partition}
Different atomic partitions can be used to compute the overlaps in Eq.~\ref{eq:sij}, however, some
partitions produce completely wrong results, such as indicating benzene as the least aromatic species of a
large set of six-membered rings.~\cite{heyndrickx:11jcc} This result is in contrast with other
electronic aromaticity indices such as FLU,~\cite{matito:05jcp,matito:06jcp_err} which is 
less sensitive to the atomic
partition~\cite{matito:06jpca} (see also Table~\ref{tbl:partition}) 
because covalent bond orders do not vary much among atomic
partitions.~\cite{matito:05jpca}
In this respect, the most suitable atomic partition~\cite{matito:07fd,heyndrickx:11jcc,rodriguez:16mp}
explored thus far is Bader's quantum theory of 
atoms in molecules (QTAIM) partition.~\cite{bader:90book} Three-dimensional atomic partitions such
as QTAIM's carry some errors due to the fact that numerical integrations of the AOM
are performed in non-regular three-dimensional atomic basins. 
In this regard, one could advocate for the use of Hilbert-space partitions that 
involve analytical integrations for 
atom-centered basis sets. 
However, it is well-known that Hilbert-space-based population analyses 
suffer from spurious results when diffuse functions, lacking a prominent atomic character,
are included in the basis sets. 
The results collected in Table~\ref{tbl:partition} suggest that 
multicenter indices suffer to a greater extent from basis-set dependency than its two-center
counterparts. Both \iring$\,$ and MCI usually produce bogus results
for basis sets containing diffuse functions if Mulliken or L\"owdin partitions
are used to define the atoms in the molecule. Mulliken results are unusually large
while L\"owdin partition values are too small. The PDI,~\cite{poater:03cej}
which measures the \textit{para}-related 2c-ESI in six member rings, is also largely affected
by the inclusion of diffuse functions, whereas FLU,~\cite{matito:05jcp}
which is constructed from the 2c-ESIs between
bonded atoms, only shows important changes for pyridine calculated with Mulliken's partition.
Interestingly, for cc-pVDZ basis set, the agreement between L\"owdin and QTAIM
partition is quite good for all the indices.
It is thus apparent that Hilbert
space based partitions can only be used in the absence of diffuse functions.
Since many chemical systems exhibit highly delocalized electronic
structures that call for the use of diffuse functions, we recommend the use of three-dimensional
space partitions for the calculation of multicenter indices.
\begin{table}[h]
\small
  \caption{\ MCI and \iring$\,$ values for benzene, cyclohexane and pyridine using QTAIM, L\"owdin
and Mulliken partitions along with cc-pVDZ (PF) and aug-cc-pVDZ (DPF) basis sets.}
  \label{tbl:partition}
  \begin{tabular*}{0.5\textwidth}{lcccccc}
    \hline
    & \multicolumn{2}{c}{QTAIM} & \multicolumn{2}{c}{Mulliken} & \multicolumn{2}{c}{L\"owdin} \\ 
    \hline
\iring    & PF & DPF & PF & DPF & PF & DPF\\
    \hline
benzene & 0.049 & 0.047 & 0.077 & 0.482 & 0.053  & 0.004\\
cyclohexane & 0.000 & 0.000 & 0.000 & 0.001 & 0.001 & 0.001 \\
pyridine & 0.046 & 0.045 & 0.076 & 0.041 & 0.054 & 0.005 \\
    \hline
MCI & PF & DPF & PF & DPF & PF & DPF\\
    \hline
benzene & 0.073 & 0.071 & 0.115 & 0.759 & 0.080  & 0.020\\
cyclohexane & 0.000 & 0.000 & 0.000 & 0.011 & 0.001 & 0.001 \\
pyridine & 0.068 & 0.066 & 0.112 & 0.254 & 0.082 & 0.021\\
    \hline
FLU & PF & DPF & PF & DPF & PF & DPF\\
    \hline
benzene & 0.000 & 0.000 & 0.000 & 0.000 & 0.000  & 0.000\\
cyclohexane & 0.096 & 0.095 & 0.067 & 0.027 & 0.083 & 0.047 \\
pyridine & 0.004 & 0.005 & 0.000 & 0.568 & 0.000 & 0.004\\
    \hline
PDI & PF & DPF & PF & DPF & PF & DPF\\
    \hline
benzene & 0.103 & 0.103 & 0.085 & 0.308 & 0.106  & 0.155\\
cyclohexane & 0.009 & 0.009 & -0.003 & 0.056 & 0.009 & 0.052\\
pyridine & 0.103 & 0.103 & 0.085 & -0.034 & 0.110 & 0.156\\
    \hline
  \end{tabular*}
\end{table}

\subsection{Computational Cost and Numerical Precision}
The operation $\mathcal{P(A)}$ in Eq.~\ref{eq:MCI} produces $n!$ summations 
and thus constitutes the bottleneck of the MCI calculation. Overall,
if carried out naively, the computational cost of MCI goes like $m^3n!/2$, where $m$ is
the number of occupied orbitals and we have taken into account that there is only $(n-1)!/2$
distinct permutations for the elements in $\mathcal{A}$.\newline

The computation of three-dimensional atomic partitions carries numerical precision
errors in the calculation of the AOMs and the total MCI error grows with the ring size. 
There is three effects as $\Delta n$ members are included in $\mathcal{A}$: 
(i) on one hand, 
the number of summations in Eq.~\ref{eq:MCI} increases
by roughly a factor of $n^{\Delta n}$, (ii) each term in the summation of Eq.~\ref{eq:iring} 
involves $\Delta n$ additional multiplications and (iii) the enlargement of the ring size
probably carries the increment of the dimension of the AOM, as larger molecules 
have more electrons and, therefore, more orbitals will be involved
in the calculation.
AOM elements
are much smaller than zero for the delocalized orbitals that contribute to the
MCI and hence, as the ring size increases the error in each summation can easily grow beyond
the machine precision. In addition, for calculations using correlated
wavefunctions (see next section) this situation worsens considerably. For single-determinant
methods (or Kohn-Sham DFT) the dimension of AOM is proportional to $N^2$, where $N$ is 
the number of electrons, but for correlated calculations the dimension goes like $m^2$,
where $m$ is the number of occupied orbitals.

There is a number of tips we can follow to  
improve the computational performance and also 
minimize the error in the MCI.
First, we should use very accurate numerical integrations schemes to avoid large
error propagations. Second, we should make a careful selection of the
orbitals involved in the MCI calculation. 
Bultinck et al. introduced the pseudo-$\pi$ method,~\cite{bultinck:08jmc,fowler:02cpl} which reduces
the size of $m$ by considering only the $\pi$ orbitals in the system. The approximation works
very efficiently for polycyclic aromatic hydrocarbons but many large
rings, such as belt-shaped M\"obius molecules, are not planar and do not render themselves to a 
classification
of its orbitals into $\sigma$ and $\pi$ types. One can also manually select the set of orbitals
that can contribute to the aromaticity of the system. Usually only valence orbitals are 
delocalized enough to contribute to the MCI and, therefore, the cost can be safely reduced
by only including these orbitals.
For highly symmetric molecules, the use of symmetry constraints can also reduce the cost by
considering only symmetrically distinct permutations.
Besides, the implementation
of Strassen algorithm for matrix multiplication can also contribute to lower the cost 
to $m^{2.8}(n-1)!/2$.\newline

Overall, the large error propagation cannot be avoided and it is an inherent
problem for the calculation of MCI and \iring; 
fortunately, these problems only show for large rings ($n\ge 10$).\newline

\subsection{Correlated wavefunctions}
The calculation of the MCI from a correlated wavefunction through Eq.~\ref{eq:ncESI}
bears a large computational cost, associated to the computation of the $n$-density.
The calculation of $n$-order densities using correlated wavefunctions for $n>3$
is typically beyond the capabilities of current computational resources, except for
very small molecules. For this reason,
many authors,~\cite{lain:04jpca} including ourselves,~\cite{cioslowski:07jpca}
have decided for approximations to the $n$-order densities that provide cost-efficient
MCI calculations using correlated wavefunctions.
Lately, we have introduced two approximations of $n$c-ESI that can be casted 
in the following expression for the MCI:
\begin{eqnarray}\label{eq:allMCI}\nonumber
\text{MCI}(\mathcal{A})=\frac{2^{(n-2)}}{n}
\sum_{\mathcal{P(A)}}\sum_{i_1\ldots i_n} \left(n_{i_1}\cdots n_{i_n}\right)^{a}
S_{i_1i_2}(A_1)\cdots S_{i_ni_1}(A_n)
\end{eqnarray}
where $a$ is a constant and $\{n_i\}$ are natural orbital occupancies. 
$a=1$ corresponds to the single-determinant expression of the $n$-densities,
whereas $a=1/2$ and $a=1/3$ are our proposals. The latter has been shown to
provide the most accurate results for 3$c$-ESIs, even better than more sophisticated
$3$-density approximations.~\cite{feixas:14jctc} However, the exact calculations 
of 3c-ESIs tend to provide very small values and certain capabilities of the
3c-ESIs are lost upon inclusion of electron correlation.~\cite{feixas:14jctc} 
For instance, Eq.~\ref{eq:allMCI} with $a=1$ is the only variant of MCI calculation
that permits
to distinguish between 3c-2e and 3c-4e interactions.~\cite{ponec:97jpca,feixas:15ctc}
For this reason, in recent aromaticity studies using correlated wavefunctions 
we have preferred the $a=1$ approximation, which provides more sensible
numbers.~\cite{cioslowski:07jpca,mercero:15cej}

\section{A new algorithm for MCI}

In order to bring down the computational cost it is much more  
profitable to reduce the number of distinct permutations in $\mathcal{A}$
than reducing the scaling of the number of basis functions. 
To this aim, we devise an
algorithm that screens the \textit{superatomic} 
overlap matrices (SAOMs hereafter) that result from the partial summation of some
indices, \eg,
\begin{eqnarray}\nonumber
S_{ij}(AB)  = \sum_k S_{ik}(A)S_{kj}(B) \\
S_{ij}(ABC) = \sum_{kl} S_{ik}(A)S_{kl}(B)S_{lj}(C) = \sum_l S_{il}(AB)S_{lj}(C) \nonumber \\
S_{ij}(ABCD) = \sum_{kl} S_{ik}(AB)S_{kj}(BC)
\label{eq:newalg}
\end{eqnarray}

These matrices can be precomputed only for the SAOMs that produce some
significant (above some threshold) value. The reduction of the number of SAOMs
decreases the cost of the MCI calculation by decreasing the number of possible
permutations.
The most efficient way to construct these matrices is by successive duplication of the
SAOM order following the binary representation of the number of atoms in the ring ($n$).
First of all, we construct a supermatrix of order $p$, where $p$ is the largest number
$2^p$ that is smaller or equal to $n$, by consecutive self-combination of lower order matrices. Namely,
we first construct the SAOMs of order 2 (2-SAOMs) from the AOMs of the atoms in our ring and the 
2-SAOMs are combined among 
themselves to build the 4-SAOMs and so on until $2^p$-SAOMs are formed. Finally, 
the pertinent low-order matrices
(generated through this process) are added to $2^p$-SAOMs until the final $n$-SAOMs are constructed. 
The MCI is calculated by summing up the traces of all the $n$-SAOMs.
Let $n$=$\sum_{i=0}^p{a_i}2^{i}$ be the binomial representation of $n$. The number of steps in the
algorithm equals $p+\sum_{i=0}^{p-1}{a_i}$.
The strengh of the algorithm is that a screening is applied to every step of the procedure
so that the number of SAOMs at each step is reduced.
The algorithm could be enhanced by the use of an orbital localization scheme to increase the 
sparsity of the SAOMs in Eq.~\ref{eq:newalg}.

\section{A new electronic aromaticity index}

Despite the speed up gained by the algorithm described in the last section, we anticipate
that the calculation of MCI on non-planar rings of more than 16 members still poses a serious
challenge. Moreover, for such large rings the numerical error in MCI and \iring$\,$ is quite big.
For these reasons, we prefer to explore the possibility to use other electronic mesures to
account for the aromaticity of large rings. 
The FLU has been successfully used in the past to study the ground-state structure of
several porphyrins.~\cite{feixas:09cjc} However, we have discussed the limitations of 
reference-based indices such as FLU to account for the aromaticity in molecules with stretched
bonds or to recognize the aromaticity of transition state structures.~\cite{matito:05the,feixas:08jcc}\newline

\begin{figure}[h]
\centering
\includegraphics[height=3cm]{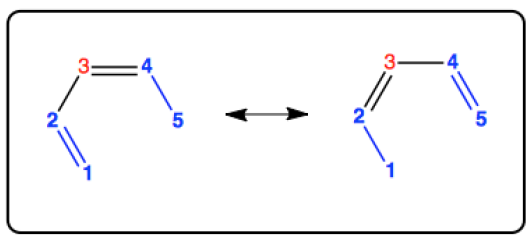}
\caption{AV1245 values for the six-membered rings.}
\label{fgr:av1245}
\end{figure}

In a previous work~\cite{feixas:10pccp} we studied the long-range delocalization patterns in
annulenes and found that the interactions separated by an even number of atoms (\eg, 1,4 and 
1,6) are larger than those separated by an odd number of atoms
(1,3 and 1,5, respectively) in aromatic molecules. 
This finding is in line with the well-known fact that the
\textit{meta}-interaction is smaller than the \textit{para}-interaction in benzene~\cite{fulton:93jpc}
(and most aromatic six-membered rings~\cite{poater:03cej}) despite the larger separation
of the atoms in \textit{para} position. Since we want to capture multicenter delocalization
effects, we suggest to average all the 4c-ESI values along the ring that keep a positional
relationship of 1,2,4,5, as indicated in Figure~\ref{fgr:av1245}. This quantity (hereafter
named AV1245), will be large if there is an important delocalization between bonds 1-2 and
4-5, a condition that only occurs in conjugated circuits. Notice that the atom in position
3 is skipped intentionally because we want to include two 1-to-4 interactions (1,4 and 2,5)
and measure the extent of transferability of the delocalized electrons in the 1-2 bond to
bond 4-5 and viceversa. We have explored other possibilities including these interactions
(\eg$\,$ AV124) but none have given as good results as this one.

AV1245 does not rely on reference values, and it does not present any limitation on 
the nature of atoms, the molecular geometry or the level of calculation (we
suggest to use Eq.~\ref{eq:ncESI} with $a=1$). Besides, it does not suffer from large
numerical precision errors, size-extensivity problems or unfavorable computational
scaling. AV1245 bears a very modest cost of $12m^3n$ that thus only grows linearly
with the ring size.
The only limitation of AV1245
is that it cannot be applied to rings of less than six members. However, for such small
rings we can safely use MCI.

\section{Computational Details}
We have studied several molecular systems that can be classified into five sets.
The first set contains a series of neutral annulenes: benzene, \ce{C8H8} (planar), 
annulene[10], annulene[12], annulene[14], annulene[16] and annulene[18],
which will be used to assess the capabilities of the indices to recognize
aromaticity as the number of pi electrons increases.
The second set contains a series of molecules that are used to validate
the performance of the new index in very different carbon-carbon bonds:
cyclohexane, cyclohexene, benzene and the transition
state of the Diels-Alder reaction occurring between butadiene and acetylene.
A series of heteroaromatic compounds with six-membered rings (6-MR) will
be used to guarantee a correct description of molecules containing 
heteroatoms: pyridine,
pyridazine, pyrimidine, pyrazine, triazine, borazine and purine.
The fourth set comprises some polycyclic
aromatic hydrocarbons: naphthalene, anthracene, phenanthrene, cyclobutadiene,
azulene and pentalene for which we
analyze the 6-MRs and the peripheral rings.
For the sake of completeness we have included two non-aromatic seven and eight-membered
rings molecules (\ce{C7H14} and \ce{C8H16}), an aromatic seven-membered ring
(\ce{C7H7+}) and hexaethynyl-carbo-benzene (\ce{C18H6}) as the molecule with a very
large ring.~\cite{lepetit:01njc}
This gives a total of seventeen 6-MRs, three 7-MRs, four 8-MRs, one 9-MR,
four 10-MRs and one 12-MR (total 30 points) that will be used to compare the
performance of MCI and AV1245. In addition, there is four 14-MRs,
one 16-MR and two 18-MRs data for which only AV1245 values will be provided.\newline

The calculations have been performed at the B3LYP/aug-cc-pVDZ level using
the g09 package.~\cite{g09} 
The AOM calculations use the QTAIM partition and have been computed with AIMall
package~\cite{aimall} and inputed into ESI-3D code.~\cite{esi3d} 

\section{Results}

\begin{figure}[h]
\centering
\includegraphics[height=6cm]{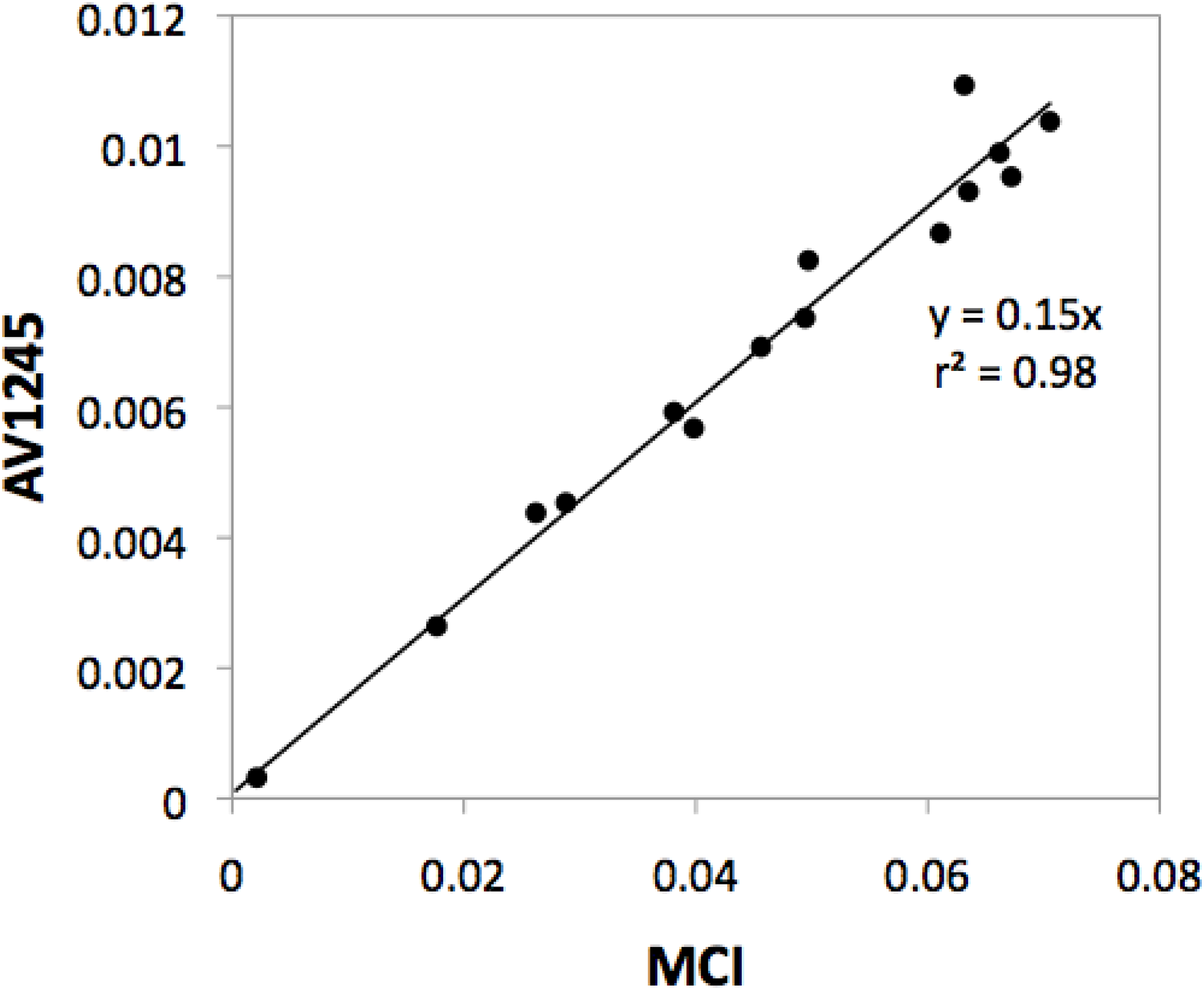}
\caption{AV1245 values for the six-membered rings.}
\label{fgr:av1245res}
\end{figure}

In Figure~\ref{fgr:av1245res} we plot AV1245 against MCI for the 6-MR studied in this 
work. The data comprises
aromatic and non-aromatic rings, the transition state of the Diels-Alder reaction (which
is a difficult case for several aromaticity indices)~\cite{matito:05the} and
the set of heteroaromatic molecules. AV1245 gives an excellent linear correlation 
with zero intercept and a Pearson coefficient of $r^2=0.98$. The only small discrepancy
between both indices is pyrazine which according to AV1245 is slightly more
aromatic than benzene.

\begin{figure}[h]
\centering
\includegraphics[height=4.5cm]{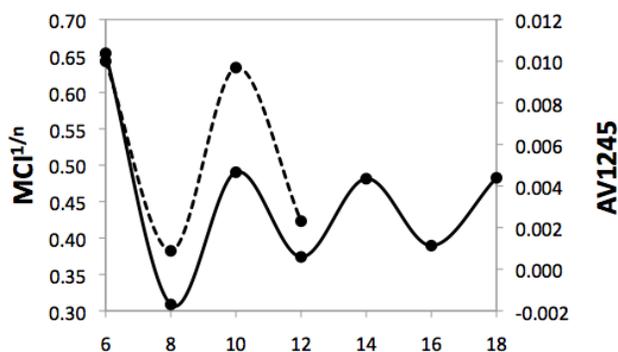}
\caption{MCI$^{1/n}$ (dashed line) and AV1245 (solid line) for a series of annulenes C$_{2m}$H$_{2m}$ $(m=3-9)$ as a function of the number of carbons.}
\label{fgr:annulenes}
\end{figure}

In Figure~\ref{fgr:annulenes} we plot AV1245 and $MCI^{1/n}$ against the number of carbon atoms 
for the series of neutral annulenes (set 1). The number of $\pi$ electrons equals the number
of carbon atoms and, therefore, according to H\"uckel's rule, one expects a zig-zag plot where
aromatic and antiaromatic compounds alternate. Although some antiaromatic compounds present 
negative MCI values, the experience indicates that many antiaromatic compounds simply give very
small number and are barely distinguishable from non-aromatic compounds.~\cite{cioslowski:07jpca}

\begin{figure}[h]
\centering
\includegraphics[height=6cm]{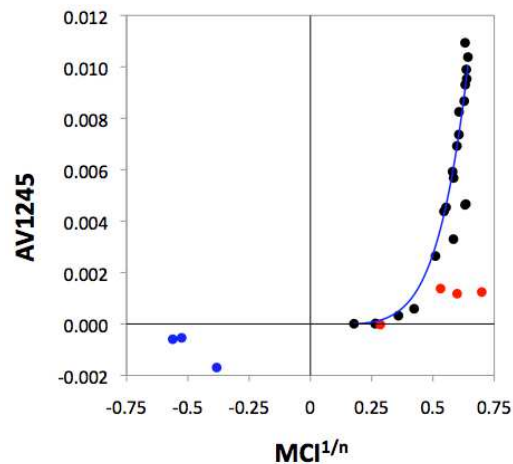}
\caption{AV1245 vs MCI$^{1/n}$ for the whole series of rings up to twelve members.
Red bullets indicate 7- and 9-membered-ring molecules and molecules with negative
MCI values are marked in blue. The solid blue line corresponds to $y=0.1$MCI$^{5.2/n}$.}
\label{fgr:av1245vsmci}
\end{figure}

Let us now analyze the ring-size dependency of both indices by including all rings from 6-MR to
12-MR in the plot of Figure~\ref{fgr:av1245vsmci}. The indices do not show a linear correlation
trend but, instead, a power-law dependency that fits most of the data points studied. A few
exceptions have been marked in blue and red. The blue bullets indicate antiaromatic
molecules that show negative values for both MCI and AV1245. One is thus deemed to conclude that,
as it happens in MCI, antiaromatic compounds exhibit either negative or very small AV1245 values. 
The red bullets correspond to 
molecules with an odd number of ring members. These molecules clearly deviate from the general
trend and put forward that the conjugation mechanisms in such rings is less obvious. The 
aromaticity in these compounds is severly underestimated by AV1245 and until further
explorations are performed AV1245 should be used in rings of even number of atoms.\newline

Finally, we collect some AV1245 values for large rings in Table~\ref{tbl:data}.
Among the polycyclic aromatic hydrocarbons (PAHs), the most aromatic 6-MR is the 
external ring of phenanthrene, whose two-ring \textit{naphathalenic} structure 
is the least aromatic 10-MR among PAHs. The peripheral ring of naphthalene
shows similar AV1245 value to the annulene[10] structure, whereas neither the peripheral
ring of anthracene or phenanthrene are as aromatic as annulene[14]. Conversely,
hexaethynyl-carbo-benzene also exhibits similar aromaticity to annulene[18], which
confirms the prominent aromatic character of this molecule.~\cite{lepetit:01njc}

\begin{table}[h]
\begin{center}
\small
  \caption{\ AV1245 values (times 1000) for selected rings.} 
  \label{tbl:data}
  \begin{tabular*}{0.3\textwidth}{rlc}
    \hline
 &   & AV1245 \\
    \hline
6-MR & benzene & 10.38 \\
6-MR & naphthalene & 5.93 \\
6-MR & anthracene ext. & 4.53 \\
6-MR & anthracene int. & 4.38 \\
6-MR & phenanthrene ext. & 6.92 \\
6-MR & phenanthrene int. & 2.64 \\
10-MR & naphthalene & 4.63 \\
10-MR & anthracene & 3.76 \\
10-MR & phenanthrene & 3.30 \\
14-MR & anthracene & 3.51 \\
14-MR & phenanthrene & 3.63 \\
10-MR & \ce{C10H10} & 4.66 \\
12-MR & \ce{C12H12} & 0.59 \\
14-MR & \ce{C14H14} & 4.35 \\
16-MR & \ce{C16H16} & 1.13 \\
18-MR & \ce{C18H18} & 4.39 \\
18-MR & \ce{C18H6}  & 4.42 \\
    \hline
  \end{tabular*}
\end{center}
\end{table}

\section{Conclusions}
We have introduced a new electronic aromaticity index, AV1245, consisting
in the average of the 4-center MCI values along the ring that keep a positional relationship 
of 1,2,4,5. AV1245 measures the extent of transferability of the delocalized electrons between
bonds 1-2 and 4-5, which is expected to be large in conjugated circuits and, therefore, in
aromatic molecules. AV1245 analyzes the aromaticity of individual rings
(local aromaticity) but it could also be used to measure the extent of all 1,2,4,5 conjugation
interactions in a molecule and, therefore, account for the global aromaticity of 
annulated species such as benzenoid macrocycles.~\cite{matito:07the}
A new algorithm for the calculation of MCI for medium-sized rings has been introduced and used to
produce the data for the calibration of the new aromaticity index.\newline

Our results indicate that AV1245 correlates very well with MCI excepting for rings with
an odd number of members. 
AV1245 does not rely on reference values, does not suffer from large numerical precision errors,
and it does not present any limitation on
the nature of atoms, the molecular geometry or the level of calculation.
It is a size-extensive measure with a small computational cost that grows linearly with
the number of ring members. Therefore, it is specially suitable for studying the aromaticity
of large molecular rings as those occurring in belt-shaped M\"obius structures or porphyrins.

\section{Acknowledgements}
I am indebted to Dr. Tom\'as Rocha for his kind hospitality during
my stay in the Institute of Chemistry at the UNAM (Mexico), where this
work was partially carried out.
The author thanks Dr. Ramos-Cordoba for fruitful discussions.
This research has been funded by Spanish MINECO 
Projects No. CTQ2014-52525-P and the Basque 
Country Consolidated Group Project No. IT588-13.
The author acknowledges the computational resources 
and technical and human support provided by SGI/IZO-SGIker UPV/EHU.

\footnotesize{
\bibliography{gen} 
\bibliographystyle{rsc} 
}

\end{document}